# NUMERICAL SIMULATIONS OF CARDIOVASCULAR DISEASES AND GLOBAL MATTER TRANSPORT

**Sergey S. Simakov** [1)], **Alexander S. Kholodov** [2)], **Alexey V. Evdokimov** [3)], **Yaroslav A. Kholodov** [4)]

1-4) Department of Applied Mathematics, Moscow Institute of Physics and Technology
9, Instituskii Lane, Dolgoprudny, 141700 Russia
1) simakov@crec.mipt.ru
2) xolod@crec.mipt.ru
3) evdokimov@crec.mipt.ru
4) kholodov@crec.mipt.ru

*Numerical model of the peripheral circulation and dynamical model of the large vessels and the heart are discussed in this paper. They combined together into the global model of blood circulation. Some results of numerical simulations concerning matter transport through the human organism and heart diseases are represented in the end.*

## INTRODUCTION

Certainly blood flow is of great importance for many processes taking place in every living organism. Particularly the processes associated with matter transport are of great interest for many applications in physiology and medicine. There are a lot opportunities for the substance penetration into circulation: inspiration, metabolic activity, endocrine control, intestinal absorption, medicine injection and others. Depending on the properties different matter may cause different influence on the organism. In this connection one of the most significant and rarely considered in mathematical modeling question is how local changes in matter concentration may cause non-local impacts.

Another widely spread class of diseases connected with different heart disorders. Especially important are stenoses of the valves and defects of inter-ventricular and inter-auricle partitions. One approach for numerical simulations of such disorders is considered in this paper.

In this paper all parts of the circulatory system are considered and joined together into one global multi-component model. This model gives adequate and detailed enough description for the blood flow distribution in different parts of circulatory system: heart, large arteries and veins of pulmonary and systemic circulation, microcirculation in tissues. On the other hand it requires a few computational resources and allows to determine basic parameters in easy and convenient manner. This allows us to simulate several processes important for medicine applications.

## 1. DYNAMICAL NET MODEL OF BLOOD FLOW IN LARGE VESSELS

Large and medium parts of pulmonary and systemic circulations have complex anatomical structure representing hierarchical irregular system of branching vessels. This structure can be described in terms of the graph theory if we assume that every vessel corresponds to the branch and every bifurcation to the node. In addition several specific nodes must be considered corresponding to the organs tissues and to the heart. Therefore blood flow going through the living organism is reduced to the flow through such graph. In fact, it is much more convenient to consider four separate graphs corresponding to the arterial and venous parts of the pulmonary and systemic circulations that are connected together by mentioned specific nodes (heart and organs). Strictly speaking these graphs can not be considered as trees since there are a lot of anastomoses especially in arterial parts. Nevertheless in many cases vessels (especially large arteries and veins) have tree-like structure.

Physically blood flow through the large and medium vessels can be regarded as pulsate flow of incompressible fluid streaming through the system of elastic tubes. So the base model for such flow is pseudo one-dimensional hydraulic model of non-stationary incompressible fluid streaming through the deformable vessel. Then it must be generalized for the case of the hierarchical branching tree-like system of vessels using appropriate boundary conditions.

For every vessel of $k^{th}$ generation we know mass and momentum conservation laws (see [1–3]):

$$\partial S_k / \partial t + \partial (u_k S_k) / \partial x = \varphi_k (t, x, S_k, u_k, r_i) \qquad (1)$$

$$\partial u_k / \partial t + \partial (u_k^2 / 2 + p_k / \rho_k) / \partial x = \psi_k (t, x, S_k, u_k, r_i) \qquad (2)$$

To close the system of equations (1, 2) it is necessary to take into account the equation of state describing the variation of the vessel cross-section depending on the transmural pressure:

$$p_k(t,x) - p_*(t,x) = \rho_k c_{k0}^2 f_k(S_k(t,x)) \qquad (3)$$

where $t$ — time; $x$ — distance along a vessel; $d_k$ — average vessel diameter; $\rho_k = const$ — density; $c_k(t,x)$ — velocity of a pulse wave; ($c_{k0}(t,x)$ — propagation speed of the small disturbances); $p_k(t,x)$ — pressure inside a vessel (counting off from the atmospheric pressure); $p_*(t,x)$ — excessive pressure in the tissues surrounding a vessel; $Q_k(t,x) = S_k u_k$ — blood flow; $R_k$ — hydrodynamic resistance; $\varphi_k(t,x,S_k,u_k,r_i)$ — source or leakage of mass; $\psi_k(t,x,S_k,u_k,r_i)$ — external forces (gravity, friction, etc.); $r_i$ — external forces; $k = 1, 2, ..., K$ — index of vessel.

## 2. FOUR-COMPARTMENT DYNAMICAL HEART MODEL

The last important part of the cardiovascular system that must be considered is the heart. The fact is that heart consists of four extensible connected chambers having inertia and hydraulic resistance. We propose to substitute chambers by extensible spheres and connecting holes by short rigid channels as represented in fig. 1. At the moment when some of the valves are closed the corresponded blood flow is zero. Some chambers may have additional connecting channels. They allow us to simulate several heart pathologies such as inter-ventricular and inter-atrial partition defects. At the entrances to the chambers 3 and 4 and at the exits of chambers 1 and 2 boundary conditions are specified taking into account coupling with the net model.

Heart period is divided into four parts:
a) Isovolumetric contraction of the ventricles: $Q_{51} = 0; Q_{62} = 0; Q_{14} = 0; Q_{23} = 0$
b) Ejection period: $Q_{14} = 0; Q_{23} = 0$
c) Isovolumetric relaxation of the ventricles: $Q_{51} = 0; Q_{62} = 0; Q_{14} = 0; Q_{23} = 0$
d) Filling period: $Q_{51} = 0; Q_{62} = 0$

The following assumptions must be considered when developing equations (see [2] for details):
— the time of opening and closing for all heart valves is zero;
— the hydrodynamic resistances of the inter-chamber channels as well as the heart inputs and outputs during every phase of the heart cycle are constant;
— the hydrodynamic resistances of the heart chambers are constant;
— the walls of the chambers are uniform and they can expand in all directions equally;
— the chamber contraction and relaxation processes are supposed to be the real isometric processes if they take place at constant volume.

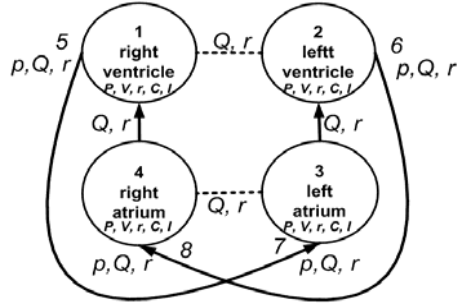

Figure 1: Four-compartment dynamical heart model considering inter-chamber partition defects.

## 3. TWO-DIMENSIONAL MODEL OF PERIPHERAL CIRCULATION

When modeling blood flow in a tissue of an organ we propose to consider isotropic region of small vessels as a plane continuum. It allows us to avoid determining a great number of individual characteristics of the small vessels and reduce the properties of these vessels to the continuously distributed parameter — permeability or tissue hydraulic resistance. Large vessels connected to the considered region can be described as small circular spots of inflow and outflow of the blood. Therefore the blood flow in the small vessels of the particular macro region of a tissue is reduced to the incompressible viscous liquid filtration through the porous medium.

In additional separate arterial and venous porous planes must be considered and diffusion of the blood among them must be taken into account. It should be pointed out that minimum grid nodes will required for approximation along this direction so it will not increase necessary computational resources. But at the same time it will allow more adequate approximation to the physiological properties of the tissues that are based on the arterial-venous blood flow and pressure difference rather than on the spatial distribution of the pressure and velocity.

Mathematical description of this model is based on the pseudo 3D Laplace operator (see [4]):

$$\Delta p_k + (p_k - p_{k+1}) r_k / R_k = -q_k r_k, \qquad (4)$$

where $k, j = 1, ..., K$ — indices of the vessel groups (arteries and veins, if $K = 2$); $p_k$ — blood pressure; $x, y$ — spatial coordinates; $q_k(x, y, p_j)$ — density of the source or the sink of blood (from and into appropriate large vessels); $r_k(x, y, p_k)$ — vascular resistance coefficient (responsible for the blood flow in the $x - y$ plane); $R_k(x, y, p_j)$ — interlayer resistance.

## 4. HEART DISORDER MODELING

The heart functioning under the normal conditions is considered in fig. 2 where results of calculations depicted by the marked curves and all other curves correspond to the experimental data [5]. Slight divergence and in the pressure (fig. 2-a) and volume (fig. 2-b) of the left heart is not significant.

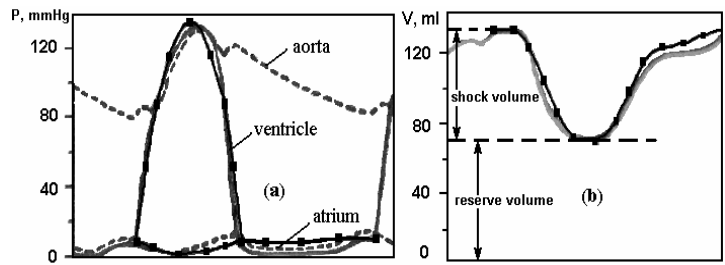

Figure 2. Parameters of the left heart.

The aortic valve stenosis and inter-ventricular partition defect are considered in the further. The stenosis of any valve could be simulated in the same manner as for aortic but this case is considered just as one of the most important. In terms of our model this phenomenon is described simply by increasing the resistance of the appropriate inter-chamber channel. The results of such simulations are shown in fig. 3 where triangles and circles depict the pressure difference before and after the aortic valve. The leftmost circle and triangle corresponds to the normal heart operation.

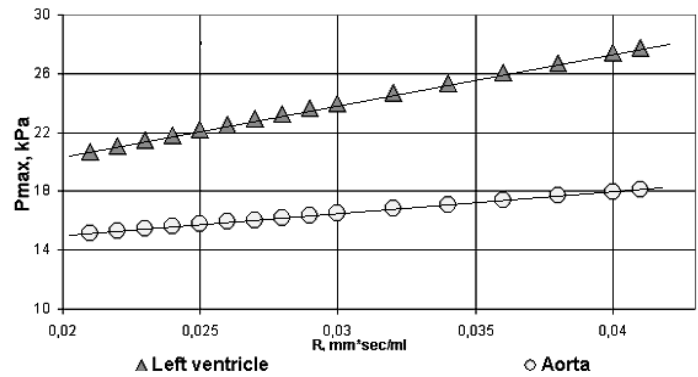

Figure 3. Aortic valve stenosis.

Simulations of the inter-ventricular and inter-atrial partition defects are described by the additional inter-chamber channels (dashed lines in fig. 1). The resistances of these channels are varied in some physiologically tolerant range. The oxygen concentration change in the venous and arterial blood is of particular interest in this case. These changes are shown in fig. 4 for the different values of the inter-ventricular partition resistance.

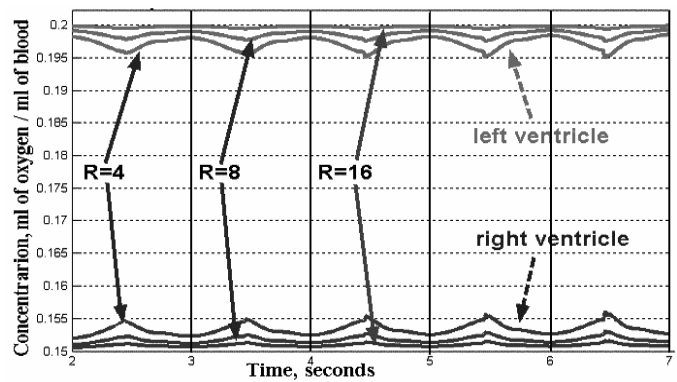

Figure 4. Inter-ventricle partition defect: arterial and venous blood mixing.
R [mm x sec/ml] — inter-ventricle partition resistance

## 5. GLOBAL MATTER TRANSPORT MODELING

The results concerning matter transport are depicted in fig. 5 and 6. The first simulation (fig. 5) reveals initial stages of the substance propagation through the organism after its inspiration in lungs. Basing on the similar idea it is possible to simulate medicine injection to the arteries or veins (fig. 6). It is supposed constant inflow of substance in the center of the artery supplying right thigh. As one may observe there is limited region that affected by this matter.

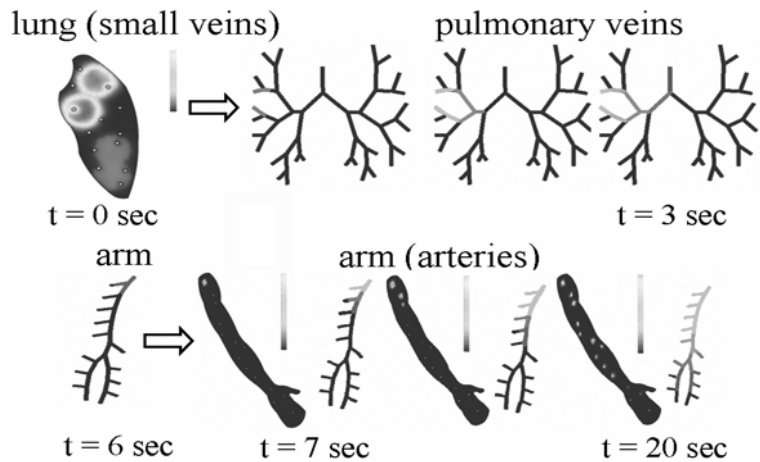

Figure 5: Initial stages of substance inspiration and propagation through the organism.

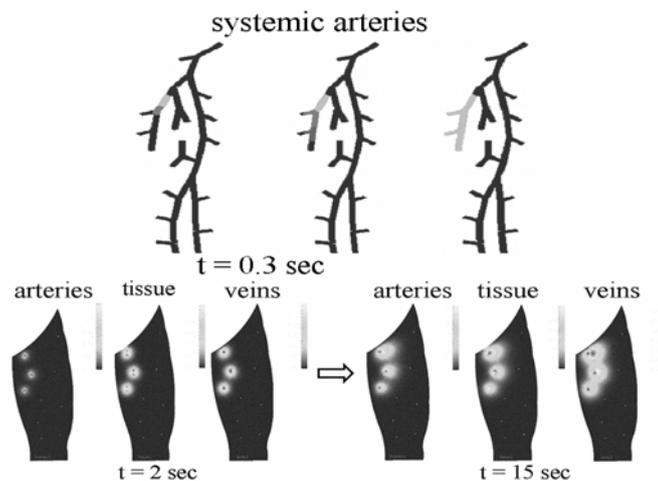

Figure 6: Medicine injection to the artery of thigh.

**CONCLUSION**

In general, global model of the blood flow in human organism was presented in this paper. Computational results show great capabilities of using such models for the tasks of global substance transfer and heart diseases. In the future the model may be improved by considering more external effects, pharmacological processes and more detailed cardiovascular graph coupled with all the most important organs that will require parallel computations.

**REFERENCES**

1. Olufsen M.S., Structured Tree Outflow Condition for Blood in Larger Systemic Arteries // Am. J. Physiol. – 1999. – №275. – P. 257-268.
2. Kholodov A.S., Some Dynamical Models of External Breathing and Blood Circulation Regarding to Their Interaction and Substances Transfer // Comp. Mod. and Med. Prog. – Moscow. – 2001. – P. 127-163.
3. Caro K., Pedley T.J., et. al. Blood Circulation Mechanics. Moscow: Mir, 1981.
4. Evdokimov A.V., Kholodov A.S., Pseudo-steady Spatially Distributed Model of Human Circulation // Comp. Mod. and Med. Prog. –Moscow. – 2001. – P. 164-193.
5. Schmidt, R.F., Thews, G., Human Physiology, Moscow: Mir, 1996, Vol. 2.